\documentclass{JHEP3}
\usepackage{graphicx}
\usepackage{amssymb}
\usepackage[mathscr]{eucal}
\usepackage{amsmath}
\usepackage{cite}
\usepackage{afterpage}
\usepackage{slashed}
\usepackage{feynmp}
\usepackage{epsfig}
\newcommand{\dd}{\mathrm{d}}

\newcommand{\gsimm}{\raise.3ex\hbox{$>$\kern-.75em\lower1ex\hbox{$\sim$}}}
\newcommand{\lsimm}{\raise.3ex\hbox{$<$\kern-.75em\lower1ex\hbox{$\sim$}}}

\newcommand{\be}{\begin{equation}}
\newcommand{\ee}{\end{equation}}
\newcommand{\ba}{\begin{eqnarray}}
\newcommand{\ea}{\end{eqnarray}}

\newcommand{\bea}{\begin{eqnarray*}}
\newcommand{\eea}{\end{eqnarray*}}

\title{Distinguishing modified gravity models}

\author{Philippe Brax\\
  Institut de Physique Th\'eorique, Universit\'e Paris-Saclay, CEA, CNRS
  F-91191Gif/Yvette Cedex, France \\ E-mail:
  \email{philippe.brax@cea.fr}}

\author{Anne-Christine Davis\\
  DAMTP, Centre for Mathematical Sciences, University of Cambridge,
  CB3 0WA, UK\\E-mail:
  \email{A.C.Davis@damtp.cam.ac.uk}}

\date{today}
\abstract{Modified gravity models with screening in local environments appear in three different guises: chameleon, K-mouflage and Vainshtein mechanisms. We propose to look for differences between these classes of models by
considering cosmological observations at low redshift. In particular, we analyse the redshift dependence of the fine structure constant and the proton to electron mass ratio in each of these scenarios.  When the absorption lines belong to
unscreened regions of space such as dwarf galaxies, a time variation would be present for chameleons. For both K-mouflage and Vainshtein mechanisms, the cosmological time variation of the scalar field is not suppressed in both unscreened and screened environments, therefore enhancing the variation of constants and their detection prospect. We also consider the time variation of the redshift of distant objects using  their spectrocopic velocities. We find that models of the K-mouflage and Vainshtein types have very different spectroscopic velocities as a function of redshift and that their differences with the $\Lambda$-CDM  template should be within reach of the future ELT- HIRES observations. }

\begin{document}

\section{Introduction}

Dark energy \cite{Astier:2012ba,Copeland:2006wr} and modified gravity \cite{Joyce:2014kja} have been thoroughly investigated since the discovery of the acceleration of the expansion of the Universe. They both involve light scalar fields on large scales which need to be screened locally where tests of gravity have been carried out\cite{Williams:2012nc}. This turns out to be achievable in just three ways for scalar field theories with second order equations of motion involving one scalar coupled conformally to matter. Chameleon \cite{Khoury:2003aq,KhouryWeltman,Brax:2004qh,Damour:1994zq,Pietroni:2005pv,Olive:2007aj,Hinterbichler:2010es,Brax:2010gi}, K-mouflage \cite{Babichev:2009ee,Brax:2012jr,Brax:2014wla}and Vainshtein \cite{Vainshtein} mechanisms depend on the properties of the local Newtonian potential or its first two derivatives respectively. The chameleon mechanism is active in regions of space where the local Newtonian potential is large enough, typically larger than $10^{-6}$ to comply with solar system tests, while for K-mouflage and Vainshtein mechanisms screening occurs inside a large radius surrounding dense objects. The Vainshtein mechanism screens all astrophysical objects such as galaxies and their clusters whereas K-mouflage does not act on galaxy clusters for instance  \cite{Brax:2015lra}. These mechanisms could be distinguished by future large scale surveys as they influence the growth of structure in rather different ways \cite{Koyama:2015vza}. Chameleons increase the growth in a scale dependent matter, with no anomalous behaviour on very large scales. K-mouflage and Vainshtein have both an effect on the background cosmology and its perturbations in a scale independent manner. Finding clear and measurable cosmological observables which could help disentangle the three mechanisms is the aim of this paper. We will not touch upon the growth of structure but focus only on small redshift observables depending on the background cosmology. We will consider both the effects of chameleon, K-mouflage and Vainshtein mechanisms on the variation of constants \cite{Carroll:1998zi,Molaro:2013saa} and on the time drift of the redshift for distant objects \cite{sandage}. In particular, a time variation of constants is only possible for chameleons in unscreened regions whereas both K-mouflage and Vainshtein do not screen cosmological time variations. As a result, a variation of constants is present even in screened regions for both K-mouflage and Vainshtein.   We will also show that the future observations by the ELT- HIRES \cite{Bonifacio:2013vfa,Pro} of the spectroscopic velocity of distant objects should shed light on both K-mouflage and Vainshtein with measurable effects in comparison with $\Lambda$-CDM whereas chameleons are almost indistinguishable.

In section 2, we discuss the variation of constants for the chameleon, K-mouflage and Vainshtein mechanisms. Chameleons with order one couplings lead to variations of the fine structure constant and the proton to electron mass ratio which is within the right ballpark for future measurements when absorbing systems lie in unscreened regions such as dwarf galaxies. In section 3, we focus on the Sandage test \cite{sandage}, i.e. the time drift of the redshift of distant objects. We show that both K-mouflage and Vainshtein could be within reach of the ELT-HIRES observations. We conclude in section 4.

\section{Variation of constants}

\subsection{Variation of $\mu$ and $\alpha$}

We will be interested in models involving one scalar field coupled to matter via a field dependent coupling function $A(\phi)$ \cite{Damour:1992we}. As a result the fundamental particles such as the electron
have a mass which depends on the scalar field in the Einstein frame\footnote{In the Jordan frame, the Planck scale is time dependent and the ratio $m_\psi/m_{\rm Pl}$ is frame independent.}.
\be
m_\psi= A(\phi) m_{\psi}^{(0)}
\ee
where $m_{\psi}^{(0)}$ is the bare mass of the particle as it would appear in the standard model Lagrangian. For such models, gauge fields are decoupled from the scalar field at tree level and a direct coupling can only appear at
the loop level \cite{Brax:2010uq}. Here we shall postulate that photons couple to the scalar with an action
\be
S_{\rm photon}=-\int d^4 x \sqrt{-g}(1+ \beta_\gamma \frac{\phi}{m_{\rm Pl}}) \frac{F^2}{4}
\ee
corresponding to the fine structure constant
\be
\alpha= \frac{\alpha_0}{1+\beta_\gamma \frac{\phi}{m_{\rm Pl}}}
\ee
where $\alpha_0\sim 1/137$ is the experimental value as measured in the laboratory and we normalise $\phi_{\rm now}=0$.
The mass of the proton is essentially due to the gluon condensate $ \Lambda_{\rm QCD}$ which can become  scalar-dependent  when the scalar couples to gluons. Here  we have the approximate mass expression \cite{Damour:1994zq}
\be
m_p = C_{\rm QCD} \Lambda_{\rm QCD} + b_u m_u + b_d m_d + C_p \alpha,
\ee
where $ C_{\rm QCD}\sim 5.2$, $b_u+b_d\sim 6$, $b_u-b_d\sim 0.5$ and $C_p \alpha_0\sim 0.63\ {\rm MeV}$. The proton to electron mass ratio $\mu$ will depend on the redshift when the scalar field becomes
dynamical and we define the variation of a quantity such as $\mu$ as $\Delta \mu= \mu (z)-\mu (0)$
\be
\frac{\Delta \mu}{\mu}= \frac{\Delta \Lambda_{\rm QCD}}{\Lambda_{\rm QCD}}-(1-\frac{b_u m_u +b_d m_d}{m_p})\frac{\Delta A}{A} +\frac{C_p \alpha_0}{m_p} \frac{\Delta \alpha}{\alpha}\sim
\frac{\Delta \Lambda_{\rm QCD}}{\Lambda_{\rm QCD}}-\frac{\Delta A}{A}
\ee
where we have used the fact that $\Delta\alpha/ \alpha$ is constrained at the $10^{-5}$ level \cite{Bonifacio:2013vfa}. Hence the  small variation of $\mu$ tests the variation of the coupling to matter $\beta$ and the dependence of the gluon condensate on the scalar is at the linear order
\be
\frac{\Delta \mu}{\mu}\sim (\beta_{\rm QCD}- \beta )\frac{\Delta \phi}{m_{\rm Pl}}
\ee
where we have $\beta= m_{\rm Pl} \frac{d\ln A}{d\phi}$ and $\beta_{\rm QCD}= m_{\rm Pl} \frac{d\ln \Lambda_{\rm QCD}}{d\phi}$. Recent observational constraints can be found in \cite{Bagdonaite:2013eia,Wendt:2013nca,Bagdonaite:2013sia}. The small variation of $\alpha$ depends on the coupling to photons
\be
\frac{\Delta \alpha}{\alpha} =- \beta_\gamma \frac{\Delta \phi}{m_{\rm Pl}}.
\ee
at the linear order too.
In the following we will evaluate these variations for the three types of screening mechanisms. The details about the cosmological dynamics for the three scenarios can be found in \cite{Brax:2004qh,Chow:2009fm,Appleby:2011aa,Brax:2014wla} and will also be recalled briefly in section 3. For each model, we consider the time variation of constants as a function of the Jordan frame redshift which corresponds to absorption line frequencies in the frame where atomic physics do not suffer from any contamination by the scalar field. The redshift in the Jordan frame is defined by
\be
1+z_J= a^{-1}_J= A (1+z_E)
\ee
where $z_E$ is the redshift in the Einstein frame where the metric reads
\be
ds^2= -dt_E^2 +a^2_E dx^2
\ee
 and the Jordan time is such that $dt_J= A dt_E$. We have normalised $A(0)=1$ and $\phi\vert_{z_J=0}=0$. The Hubble rate in the Jordan frame is given by $H_J\equiv \frac{d\ln a_J}{dt_J}$
\be
H_J=A^{-1} ( H_E+ \frac{d\ln A}{dt})
\ee
In section 3, we will also consider the time variation of the redshift of distant objects as measured using spectroscopy and therefore depending on the Hubble rate in the Jordan frame. In the following, we shall suppress the indices $E$ and $J$ as they should be clear from the context.

\subsection{ The models}

\subsubsection{Chameleons}
In this paper, we shall focus on three types of models. The first ones, chameleons, are scalar tensor-theories whose action can be written as in the Einstein frame
\be
S=\int d^4 x \sqrt{-g}(\frac{R}{16\pi G_N} -\frac{(\partial \phi)^2}{2} -V(\phi))+S_m (\psi, A^2(\phi) g_{\mu\nu})
\label{act}
\ee
where $A(\phi)$ is an arbitrary function. The coupling to matter of the scalar field is simply given by
\be
\beta (\phi)= m_{\rm Pl} \frac{d \ln A(\phi)}{d \phi}.
\ee
as we have already used.
One important feature of these models  is that the scalar field dynamics are determined by an effective potential which takes into account the presence of the conserved matter density $\rho$ of the environment
\be
V_{\rm eff}(\phi) =V(\phi) +(A(\phi)-1) \rho.
\ee
Scalar-tensor theories whose effective potential $V_{\rm eff}(\phi)$ admits a density dependent minimum $\phi (\rho)$, the chameleons,  can all be described
parametrically from the sole knowledge of the mass function $m(\rho)$ and the coupling $\beta (\rho)$ at the minimum of the potential \cite{Brax:2011aw,Brax:2012gr} using the parametric integral
\be
\frac{\phi (\rho)-\phi_c}{m_{\rm Pl}}= \frac{1}{m_{\rm Pl}^2}\int_{\rho}^{\rho_c} d\rho \frac{\beta (\rho) A(\rho)}{m^2(\rho)},
\ee
where we have identified the mass as the second derivative
$
m^2 (\rho)= \frac{d^2 V_{\rm eff}}{d\phi^2}\vert_{\phi=\phi (\rho)}
$
and the coupling
$
\beta (\rho)= \frac{d\ln A}{d\phi}\vert_{\phi=\phi(\rho)}.
$
In the following, we shall only consider models where $A(\rho)\sim 1$, $m(\rho)$ increases with $\rho$ as befitting the chameleon mechanism and $\beta (\rho)$ decreases with $\rho$ to enhance the screening property. These requirements imply that $\phi(\rho)$ is a decreasing function of $\rho$.
We will also find it more convenient to parameterise $m(\rho)$ and $\beta(\rho)$ in a simple way using the time evolution of the matter density of the Universe
$
\rho(a)=\frac{\rho_{0E}}{a^3}
$
where $a$ is the scale factor whose value now is $a_0=1$ and $\rho_{0E}=3\Omega_{0mE} H^2_{0E} m_{\rm Pl}^2$.

We will focus on two typical chameleon models. The first ones are the
large curvature $f(R)$ models \cite{Hu:2007nk} that  have the chameleon property and can be reconstructed using $\beta(a)=1/\sqrt{6}$ and the mass function
\be
m(a)= m_0 (\frac{4\Omega_{\Lambda 0E}+ \Omega_{m0E} a^{-3}}{4\Omega_{\Lambda 0E}+ \Omega_{m0E}})^{(n+2)/2}
\ee
where the mass on large cosmological scales is given by
\be
m_0= H_{0E} \sqrt{\frac{4\Omega_{\Lambda 0E}+ \Omega_{m0E} }{(n+1) f_{R_0}}},
\ee
and
$\Omega_{\Lambda 0E} \approx 0.73$, $\Omega_{m0E}\approx 0.27$ are the dark energy and matter density fractions now \cite{Brax:2012gr}. Local tests of gravity require that in the solar system
\be
f_{R_0}\lesssim \ 10^{-6}
\ee
which we will use as template throughout. Stronger bounds at the $10^{-7}$ have been obtained from the astrophysics of stars \cite{Jain:2012tn}.

The environmentally dependent dilaton \cite{Brax:2010gi}  is another type of model which is  inspired from string theory in the large string coupling limit. It
has  an exponentially runaway potential
and a quadratic coupling function $A(\phi)$.
These models can be described using the coupling function
\be
\beta(a)= \beta_0 a^3
\ee
where  $\beta_0=\frac{\Omega_{\Lambda 0E}}{\Omega_{m0E}}\sim 2.7$, and the mass function
\be
m^2(a)= 3 A_2 H^2(a)
\ee
is proportional to the Hubble rate with the mass on cosmological scales now given by $m_0=\sqrt{3 A_2} H_{0E}$. Solar system tests require that $A_2 \gtrsim 10^6$.

When the Hubble rate is normalised in the Jordan frame, all the previous formulae need to be reexpressed as a function of the Hubble rate and the matter fraction in the Jordan frame. This is made explicit below in section 3 and this is what has been used in the Figures for $f(R)$ and dilaton models. In practice, the difference between the two frames for these models is tiny.

For all these chameleon models and at
low redshift, unscreened objects are typically characterised by
\be \Phi_N \lesssim \frac{H_{0E}^2}{m_0^2}. \ee
where $m_0$ must then satisfy \cite{Brax:2011aw,Wang:2012kj}
\be
\frac{m_0}{H_{0E}} \gtrsim 10^3.
\ee
from local gravitational tests.
As a result, unscreened astrophysical objects must necessarily have a low Newtonian potential
\be
\Phi_N \lesssim 10^{-6}.
\label{weak}
\ee
For these objects such as dwarf galaxies, the particle masses and the fine structure constant at redshift $z$ would be the cosmological one and therefore observations of these regions of the sky
would give direct access to the dynamics of the chameleon mechanism on cosmological scales.

\begin{figure*}
\centering
\epsfxsize=7 cm \epsfysize=7 cm {\epsfbox{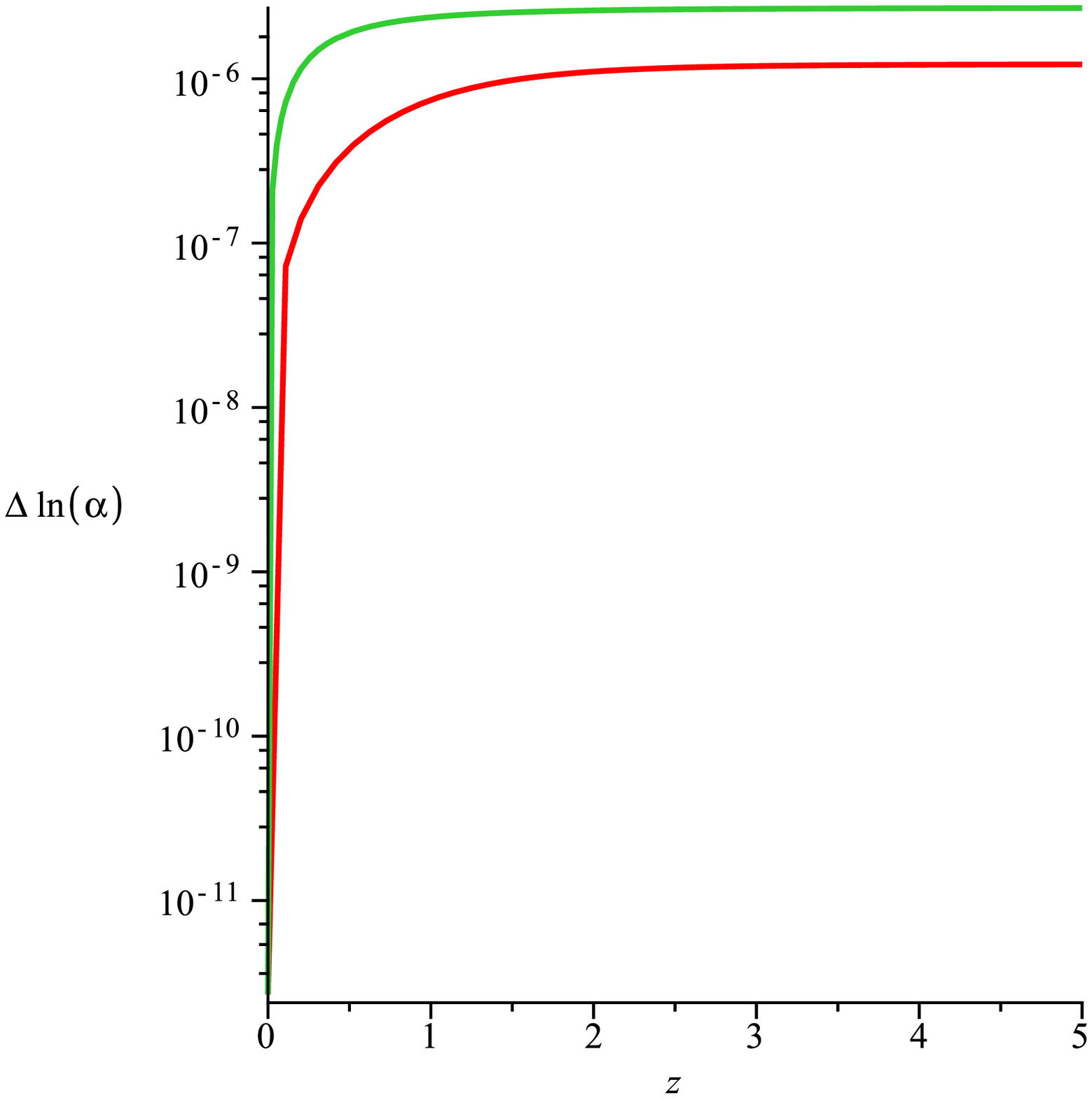}}
\epsfxsize=7 cm \epsfysize=7 cm {\epsfbox{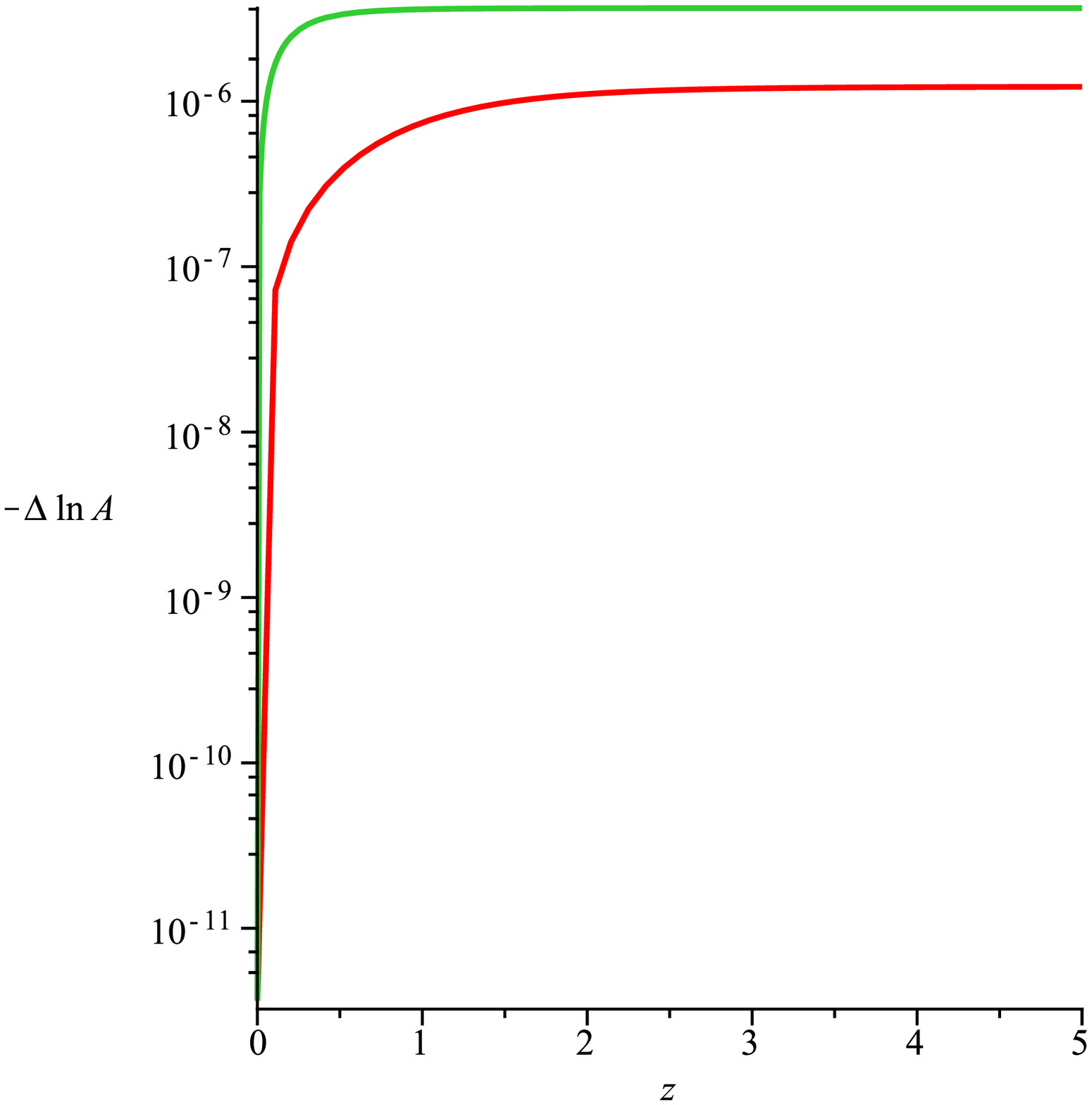}}
\caption{The variation of the fine structure constant and the coupling $\ln A(\phi)$  as a function of  redshift for dilatons (top-green) with $A_2=10^6$ and f(R) (bottom-red) with $f_{R_0}=10^{-6}$.}
\end{figure*}

We have plotted in Figure 1 the variation of the fine structure constant  and $\ln A$ for $f(R)$ and dilaton models where we have taken $f_{R_0}=10^{-6}$, $n=1$ for $f(R)$ and $A_2=10^6$ for the dilaton.
We have taken $\beta_\gamma=1$ for the variation of $\alpha$.

As can be seen, the variations of the fine structure constant in unscreened regions are of the order $10^{-6}$, i.e. comparable with the present experimental bounds. The variation of $\ln A$ is also within the $10^{-5}$ experimental bound at intermediate  redshifts for the variation of $\mu$ \cite{Levshakov:2010tj,Levshakov:2010cw}. For these models, a detection of the variation of constants in unscreened regions would be correlated with deviations of the growth of structure in the
Mpc range. For effects on much smaller scales coming from larger values of $m_0/H_0\gg 10^3$, the variation of constants would be highly suppressed and very likely unmeasurable.

\subsection{K-mouflage}

The K-mouflage mechanism can be exemplified using the
 scalar field models whose action in the Einstein frame is
\be
S  = \int \dd^4 x \; \sqrt{-g} \left( \frac{R }{16\pi G_N}+ {\cal M}^4 \, K(\chi)
\right)+S_m (\psi, A^2(\phi) g_{\mu\nu})
\ee
with the reduced kinetic term $\chi$ is defined as
\be
\chi = - \frac{1}{2{\cal M}^4} (\partial\phi)^2.
\ee
Here, ${\cal M}^4$ is an energy scale that is of the order of the current energy density  in order to recover the late-time accelerated expansion
of the Universe.
The  cosmological behaviour of a canonically normalised scalar field  together  with a cosmological constant term
$\rho_{\Lambda} = {\cal M}^4$  is recovered at late time in the weak-$\chi$ limit if we have:
\be
\chi \rightarrow 0 : \;\;\; K(\chi) \simeq -1 + \chi + ... ,
\ee
where the dots stand for higher-order terms.
For the kinetic function $K(\chi)$, we consider as in \cite{Brax:2014wla}
the polynomials
\be
K(\chi) = -1 + \chi + K_0 \, \chi^m ,
\ee
and we focus on the low-order case $m=3$ with $K_0=1$ as this model does not suffer from all the instabilities that plague K-mouflage models when $K_0<0$ or $m$ is even.

Solar system tests of gravity imply that \cite{Barreira:2015aea}
\be
\beta\le 0.1
\ee
from the time variation of Newton's constant which must satisfy  $\frac{d\ln G_N}{dt_J}\vert _{\rm now} \le 2.10^{-2} H_{0J}$ in the Jordan frame \cite{Babichev:2011iz}. Locally in the solar system Newton's constant is modified and becomes $G_N(1+ \frac{2\beta^2}{K'})$ implying that  we must have
$\chi_{s.s.} \lesssim  \chi_\star=-10^6$ and for such values of $\chi$ we must have $K\gtrsim K_\star =10^3$ to satisfy the Cassini bound \cite{Bertotti:2003rm} on fifth forces in the solar system. Even in such  screened environments, the scalar field is sensitive to the time drift of the background field on cosmological scales, i.e. $\phi(r,t) \sim \phi_{\rm cosmo}(t) + \phi_{s.s}(r)$. This implies that both the fine structure constant and the proton to electron mass ratio would vary at low redshift for K-mouflage models.

We have plotted the variations of $\alpha$ and $\ln A$ for cubic K-mouflage models with $K_0=1$, $\beta=0.1$ and $\beta_\gamma=10^{-5}$. As can be seen in Figure 2, the coupling $\beta_\gamma$ has to be that small in order to pass the current bounds on the variation of $\alpha$. The variation of $\ln A$ is too large to be compatible with the bounds on the variation $\mu$. This implies that $\delta\beta_{\rm QCD}= \beta_{\rm QCD} -\beta$ must be less than $10^{-5}$. This is the type of tuning that one has to face to make the K-mouflage scenario viable. This can be made natural when the QCD phase transition is taken to happen in the Jordan frame implying that $\beta_{\rm QCD}= \beta$.

\begin{figure*}
\centering
\epsfxsize=7 cm \epsfysize=7 cm {\epsfbox{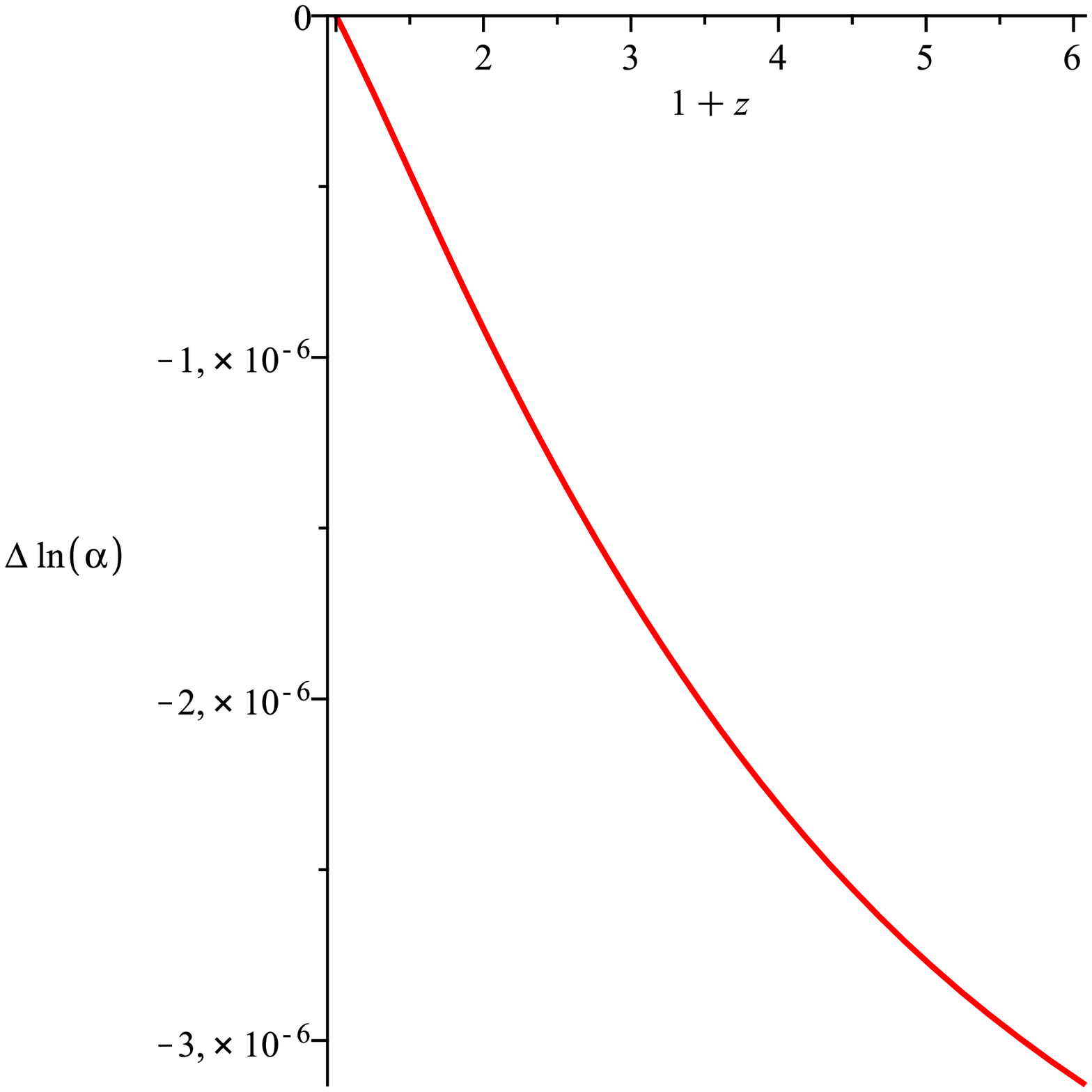}}
\epsfxsize=7 cm \epsfysize=7 cm {\epsfbox{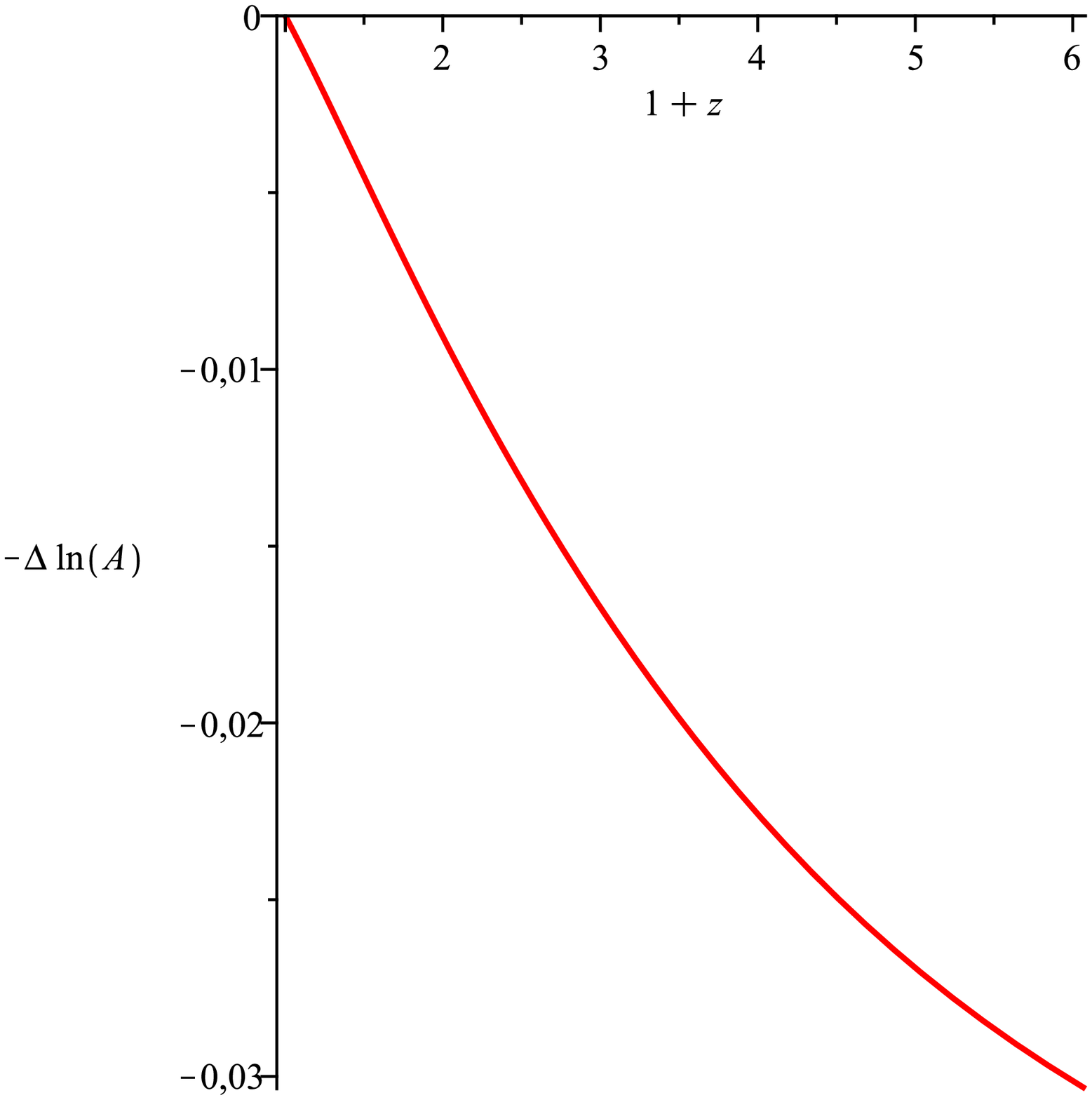}}
\caption{The variation of the fine structure constant and the coupling $\ln A(\phi)$ as a function of  redshift for the cubic K-mouflage model with $K_0=1$, $\beta=0.1$ and $\beta_\gamma=10^{-5}$.}
\end{figure*}

\subsection{Vainshtein}

We now turn to the Vainshtein mechanism which can be nicely exemplified using the Galileon models \cite{Nicolis:2008in}.   Their Lagrangian is given by the non-linear expression
\begin{equation}
\mathcal{L} = -\frac{c_2}{2}(\partial \phi)^2 -\frac{c_3}{\Lambda^3}\Box\phi (\partial \phi)^2 -\frac{c_4}{\Lambda^6}{\cal L}_4 -\frac{c_5}{\Lambda^9}{\cal L}_5 ;,\label{lag}
\end{equation}
where we focus on Galileons models with $c_2>0$ as can be derived from stable brane constructions with positive tensions \cite{deRham:2010eu}. The common scale
\be
\Lambda^3 =H_0^2 m_{\rm Pl}
\ee
is chosen to lead to dark energy in the late time Universe.

\begin{figure*}
\centering
\epsfxsize=7 cm \epsfysize=7 cm {\epsfbox{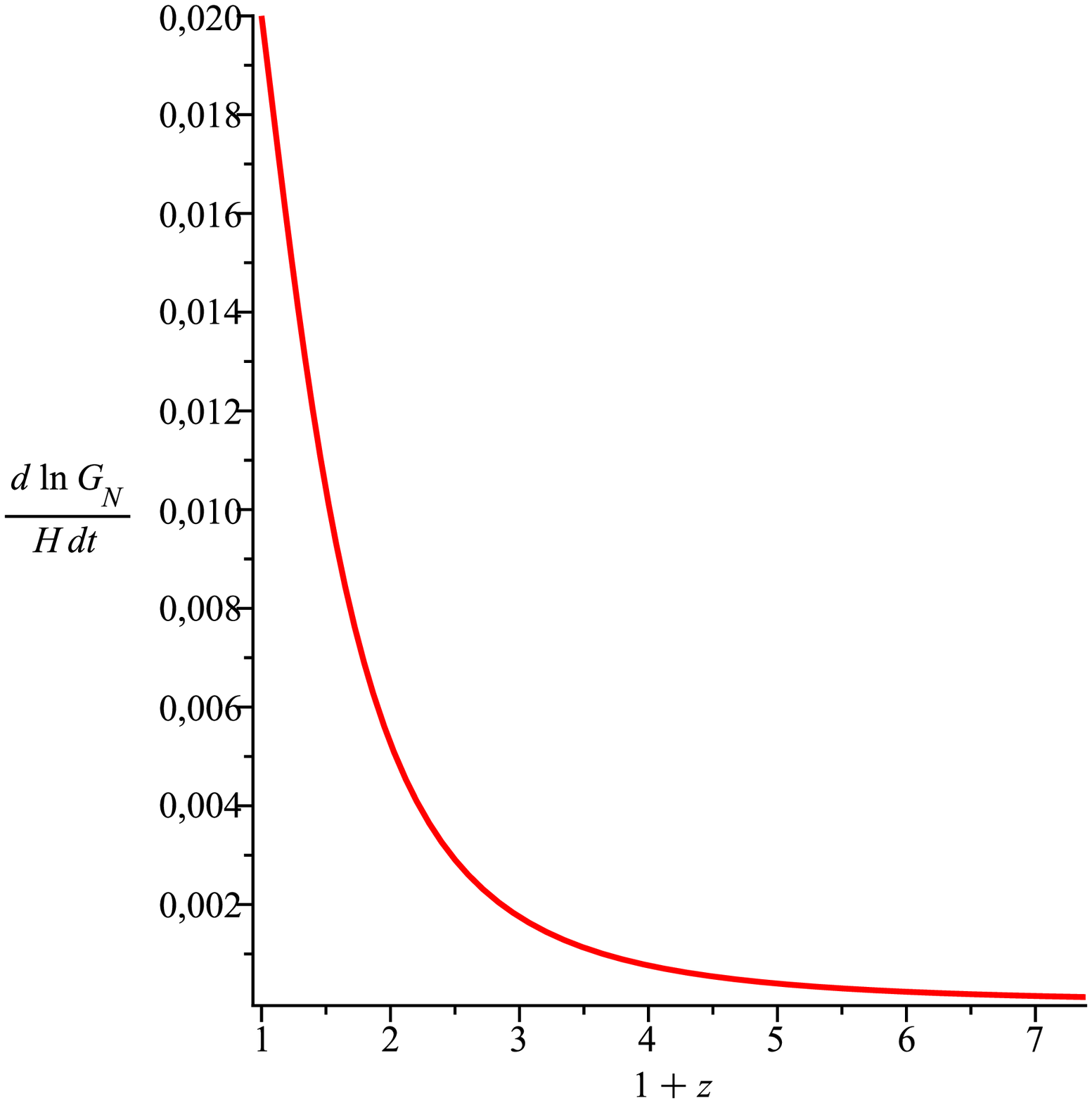}}
\epsfxsize=7 cm \epsfysize=7 cm {\epsfbox{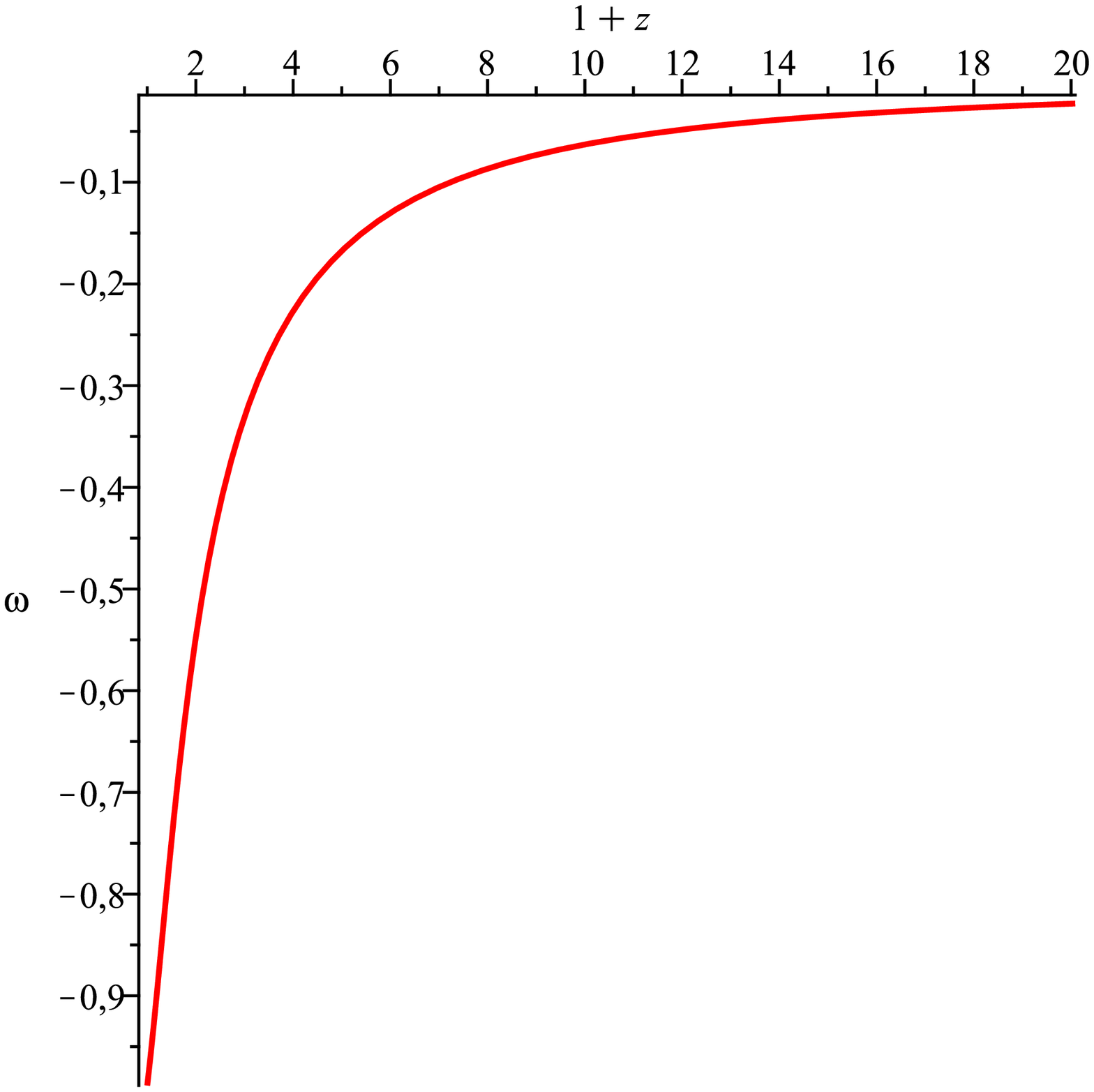}}
\caption{The time variation of Newton's constant and the effective equation of state as a function of  redshift for the quartic Galileon  model with $\bar c_2=1$, $\bar \beta_b=0.01$, $\bar\beta=0.32$  and $\bar \beta_\gamma=10^{-5}$.}
\end{figure*}

Contrary to chameleons and K-mouflage, dark matter and baryons have to couple differently to the scalar field \cite{Brax:2014vla}. Indeed, the coupling to dark matter is crucial to obtain an effective equation of state of order $-1$ in the recent past of the Universe. On the other hand, for such large values of $\beta$ and if $\beta_b=\beta$, the time variation of Newton's constant now would lead to large changes of the planetary trajectories in the solar system.
This can be remedied by taking $\beta_b< \beta$. In this case, the Newtonian constant corresponding to the Jordan frame of baryonic mass must fulfill the bound on $\frac{d\ln G_N}{Hdt_J}$. This can be achieved when
\be
\bar \beta_b \lesssim 10^{-2}
\ee
The variation of the equation of state and of Newton's constant can be seen in Figure 3.

The Galileon Lagrangian depends on
 the higher order terms  which are given by
 \begin{align}
 {\cal L}_4=&(\partial \phi)^2\left[2(\Box \phi)^2 -2 D_\mu D_\nu \phi D^\nu D^\mu \phi -R\frac{(\partial\phi)^2}{2}\right]\nonumber \\
 {\cal L}_5=& (\partial\phi)^2\left[(\Box\phi)^3 -3(\Box\phi)D_\mu D_\nu \phi D^\nu D^\mu \phi + 2 D_\mu D^\nu \phi D_\nu D^\rho\phi D_\rho D^\mu\phi\right.\\
&\left. -6 D_\mu\phi D^\mu D^\nu \phi D^\rho \phi G_{\nu\rho}\right].\nonumber
 \end{align}
and these terms play an important role cosmologically.
The Galileons in a Friedmann-Robertson-Walker background have  equations of motion in the Jordan frame which can be simplified using $x= \phi'/m_{\rm Pl}$. Their behaviour depends entirely on the rescaled couplings (see section 3) \cite{Neveu:2013mfa}
$
\bar c_i= c_i x_0^i, \ i=2\dots 5, \ \ \bar \beta= \beta x_0,\ \bar \beta_b = \beta_b x_0, \ \bar \beta_{\gamma}= \beta_\gamma x_0
$
where $x_0$ is the value of $x$ now. For these models, a non zero coupling to CDM is necessary to have an equation of state of dark energy close to -1 now. Typically, we shall take $\bar c_2=1$, $\bar c_3=1.2$ and $\bar \beta=0.32$.

\begin{figure*}
\centering
\epsfxsize=7 cm \epsfysize=7 cm {\epsfbox{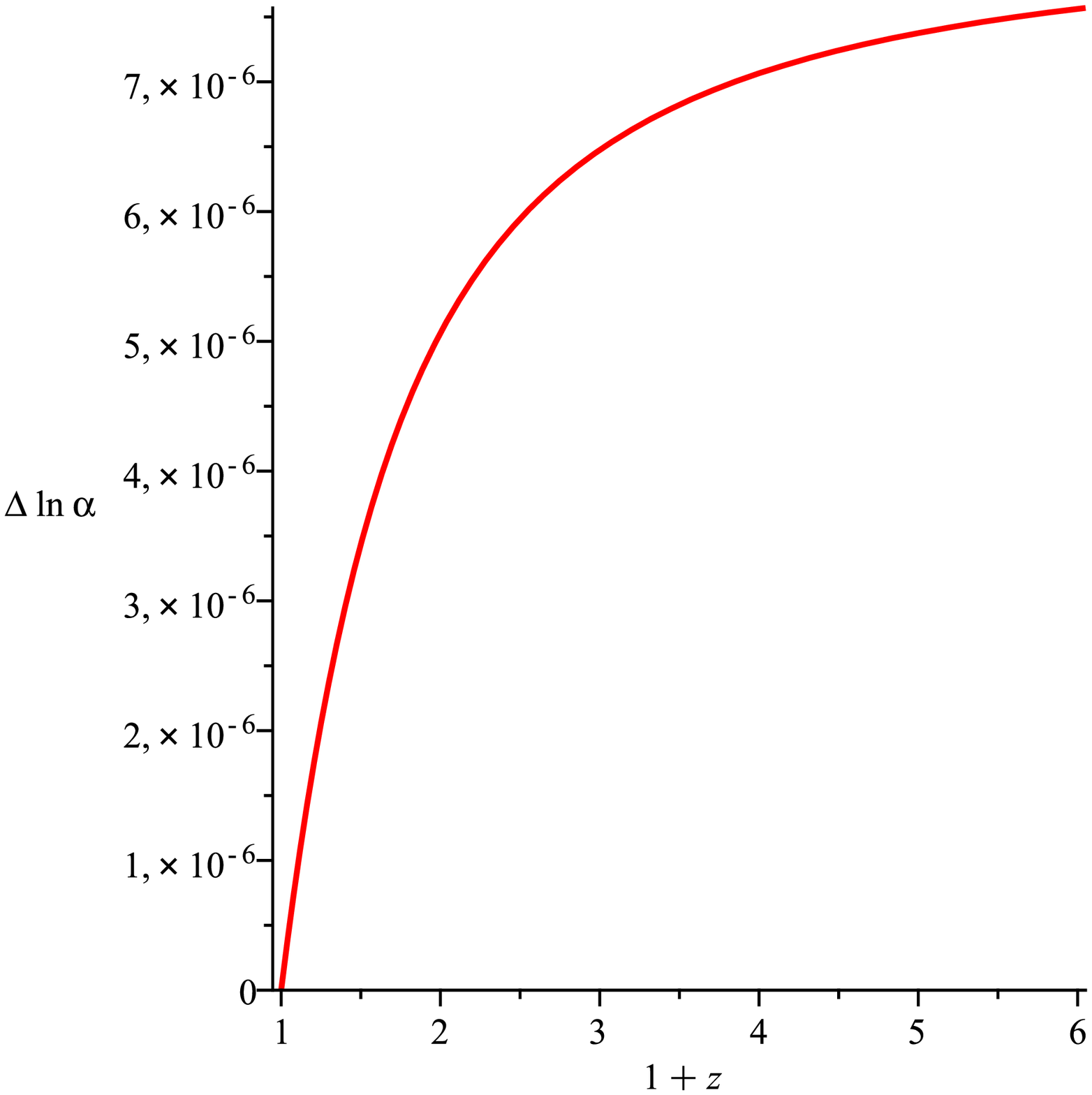}}
\epsfxsize=7 cm \epsfysize=7 cm {\epsfbox{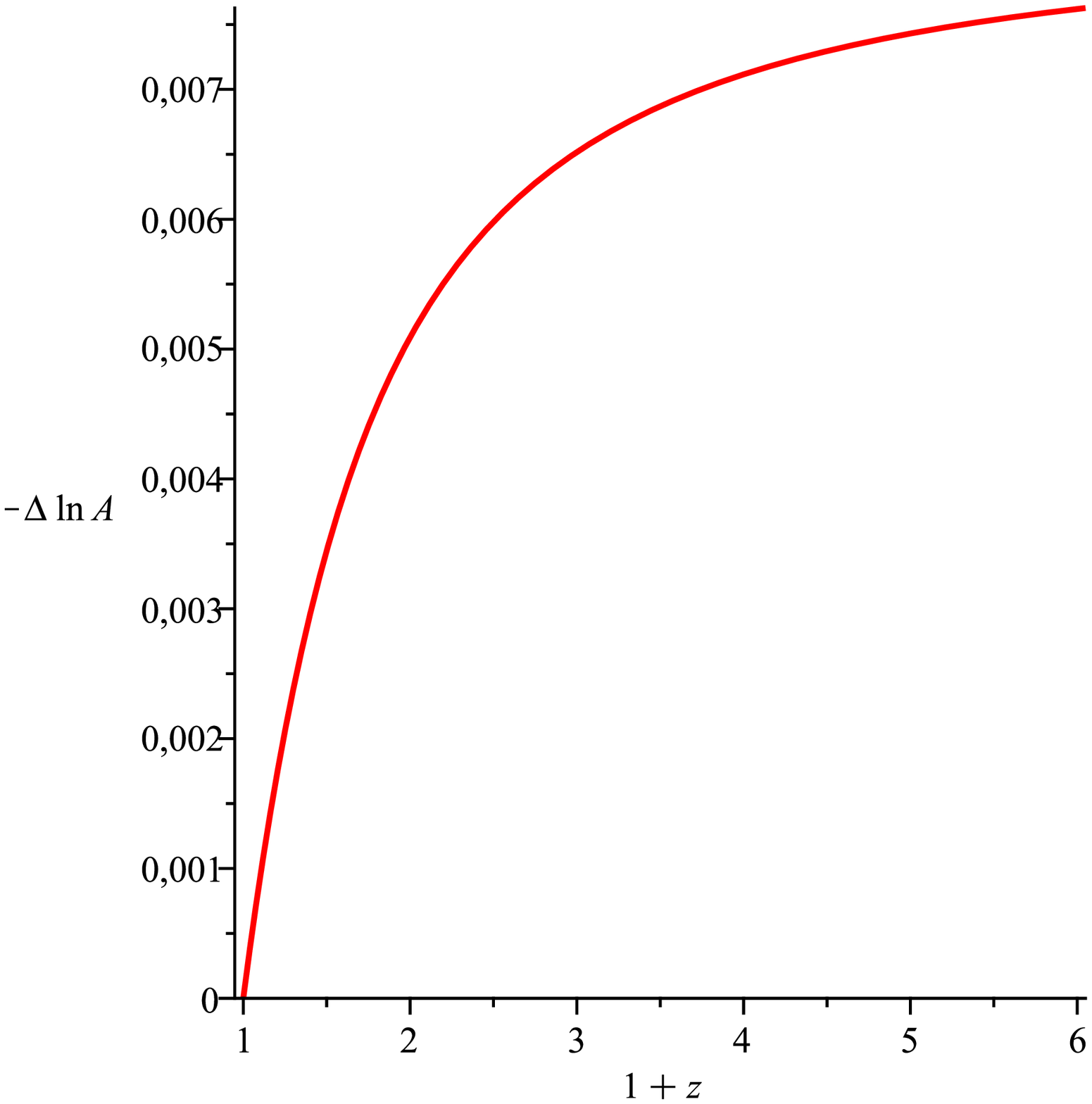}}
\caption{The variation of the fine structure constant and the coupling $\ln A(\phi)$  as a function of  redshift for the quartic Galileon  model with $\bar c_2=1$, $\beta_b=0.01$ and $\beta_\gamma=10^{-5}$.}
\end{figure*}
Again a very small value of $\bar\beta_\gamma$ is required to keep the variation of $\alpha$ within the experimental bounds. The variation of $\ln A$ is too large to comply with the bounds on $\mu$, hence  a certain degree of fine-tuning on the value of $\delta \beta_{\rm QCD}\lesssim 10^{-5}$ must be invoked in order to satisfy the current bounds. This can be made natural when the QCD phase transition is taken to happen in the Jordan frame implying that $\beta_{\rm QCD}= \beta$.

\section{Spectroscopic velocity}
\subsection{Time dependence of red-shift}
Another important effect of the modified gravity models is the time drift of the redshift measured for distant objects, the Sandage effect \cite{sandage,Loeb:1998bu,Corasaniti:2007bg,martins}. This spectroscopic velocity results from the time dependence of the Hubble rate which differs from its $\Lambda$-CDM counterpart. Interpreted as coming from the Doppler effect, the spectroscopic velocity is given by
\be
\frac{v}{c}= \frac{H_0 \Delta t}{1+z}((1+z)- \frac{H(z)}{H_0})
\ee
where $\Delta t$ is the observational time span a. The Hubble rate here is the one in the Jordan frame.
In each of the three scenarios, we will calculate the Hubble rate in the Jordan frame where Newton's constant is time dependent.

\subsection{Chameleons}
The cosmological chameleon field follows the attractor which is the minimum of the effective potential $V_{\rm eff}(\phi)$ \cite{Brax:2004qh,Brax:2011aw}. This allows us to write the Friedmann equation in the Einstein frame as
\be
H_E^2= \frac{V(a) + A(a) \frac{3\Omega_{m0E}m_{\rm Pl}^2 H_{0E}^2}{a^3}}{3m_{\rm Pl^2}(1- \frac{3}{2} \frac{\rho_m^2 \beta^2(a)}{m^4(a) m_{\rm Pl}^4})}
\ee
where we have
\be
\rho_{m}= \frac{3\Omega_{m0E} H_{0E}^2m_{\rm Pl}^2}{a^3}
\ee
and we have used $\Omega_{\Lambda0E}=(1- \frac{3}{2} \frac{\rho_{m0}^2 \beta^2_0}{m^4_0 m_{\rm Pl}^4})\Omega_{\Lambda 0}$ and $\Omega_{m0E}=(1- \frac{3}{2} \frac{\rho_{m0}^2 \beta^2_0}{m^4_0 m_{\rm Pl}^4})\Omega_{m0}$.
In practice, the correction term in the pre-factor  is very small. The coupling function can be expressed as
\be
A(a)=1-3\int_a^1 \frac{\beta^2(a') \rho_m(a')}{a'm^2(a') m_{\rm Pl}^2} da'
\ee
 and the potential term is
\be
V(a)= 3\Omega_{\Lambda 0E} H_{0E}^2 m_{\rm Pl}^2 -3\int_a^1 \frac{\beta^2(a') \rho_m^2(a')}{a'm^2(a') m_{\rm Pl}^2} da'
\ee
which can be easily evaluated for both $f(R)$ models and dilatons. The Hubble rate $H_0$ in the Jordan frame is related to the Hubble rate in the Einstein frame by
\be
H_0= H_{0E}(1+3 \frac{\beta^2_0 \rho_{m0}}{m^2_0 m_{\rm Pl}^2})
\ee
where the correction term is tiny.
In Figure 5, we have shown the evolution of the difference
\be
\frac{\Delta v}{c}= \frac{v_{\rm chameleon}- v_{\Lambda CDM}}{c}
\label{eq:dv}
\ee
between the spectroscopic velocities of the $\Lambda$-CDM case and the chameleon models ($f(R)$ and dilaton). The difference $\Delta v$ is similarly defined for K-mouflage and Galileons.
The spectroscopic velocities differ from $\Lambda$-CDM at the $10^{-5}$ level for $f(R)$ and at the $10^{-3}$ level for the dilaton. This will not be testable observationally in near future (see below).
\begin{figure*}
\centering
\epsfxsize=7 cm \epsfysize=7 cm {\epsfbox{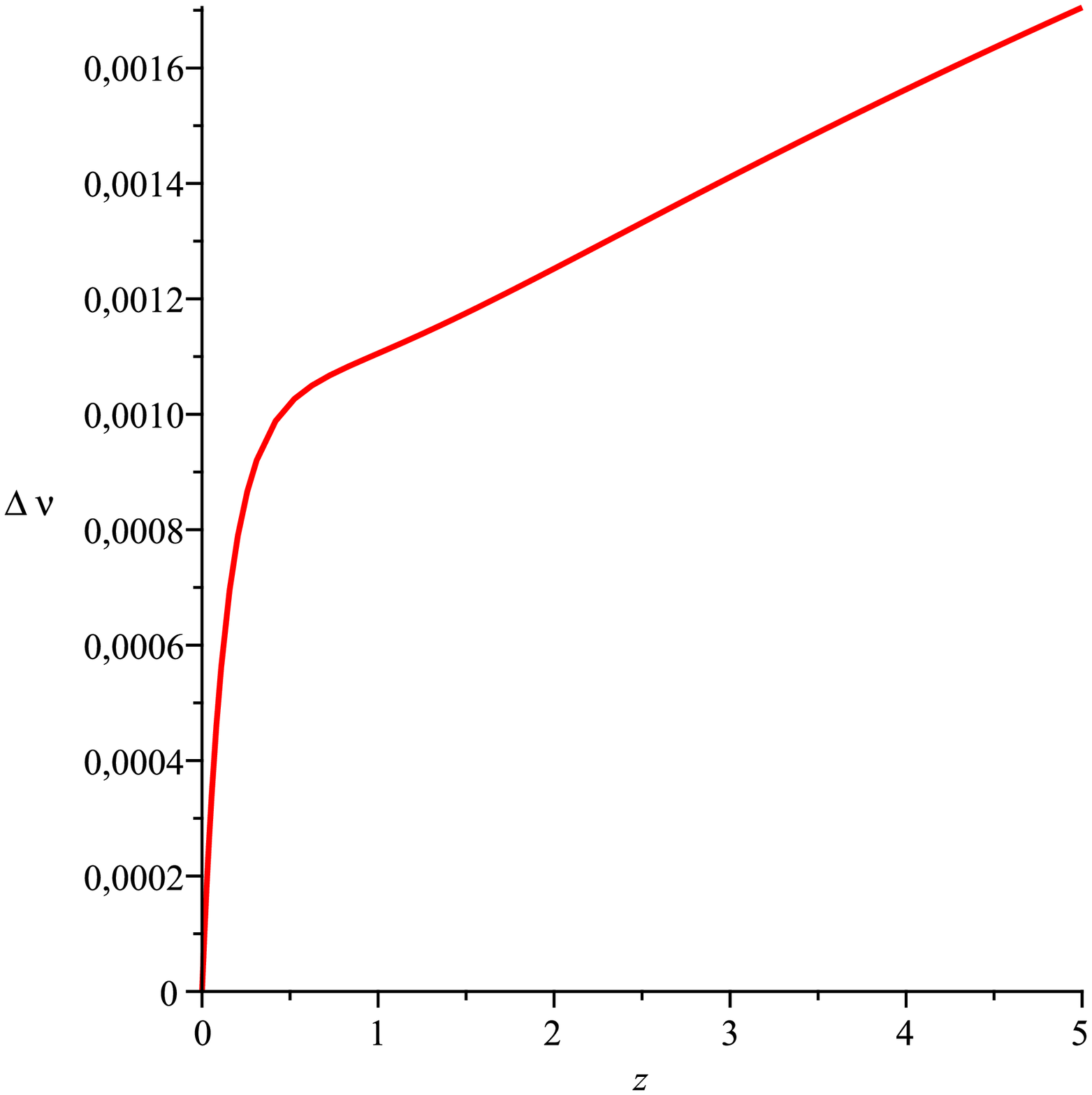}}
\epsfxsize=7 cm \epsfysize=7 cm {\epsfbox{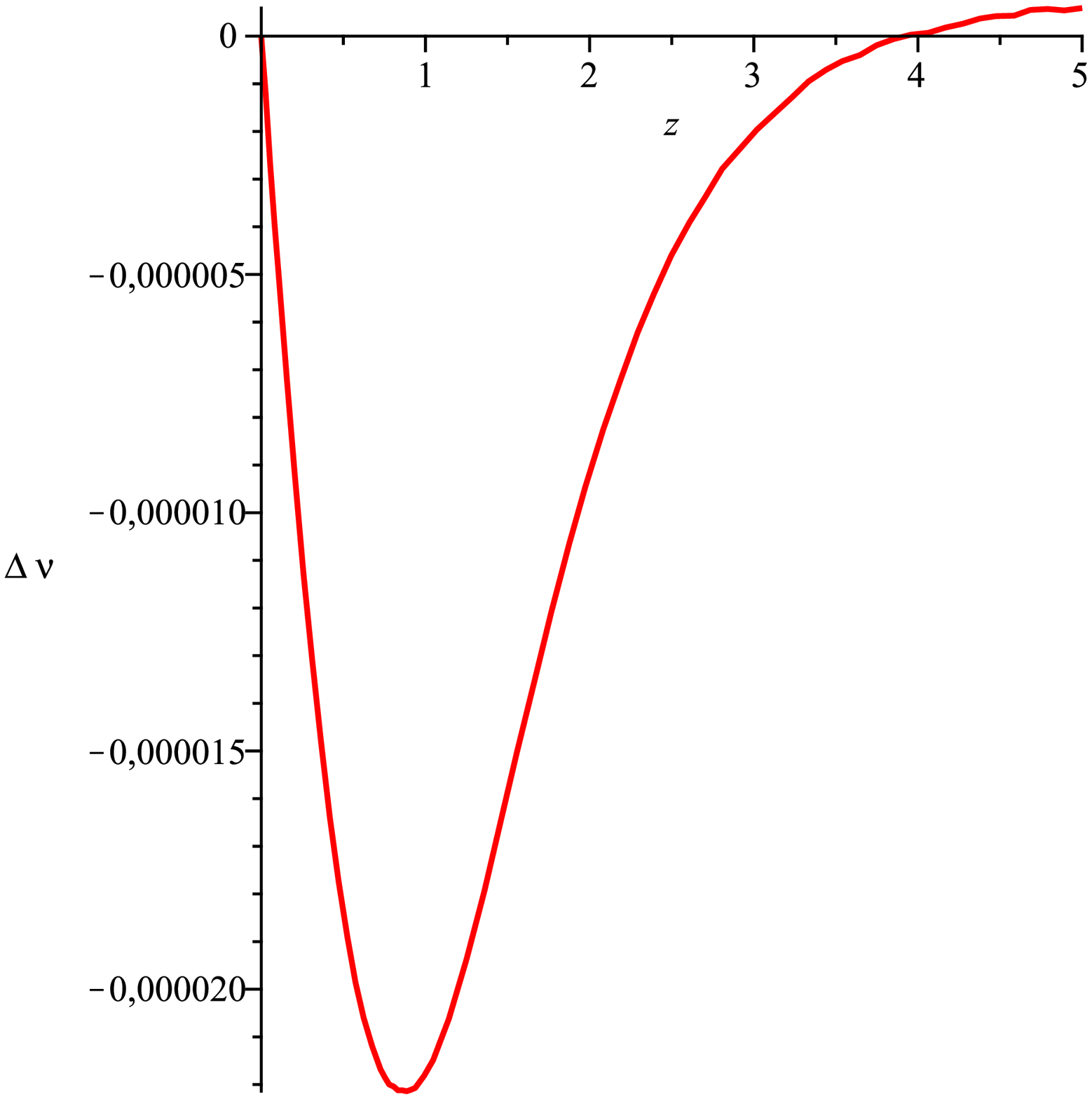}}
\caption{The variation of the spectroscopic velocities $\Delta v$  defined in  equation (3.7) in cm/s   as a function of  redshift for the dilaton (left) with $A_2=10^6$ and $f(R)$ (right)  models with $n=1$ and $f_{R_0}=10^{-6}$.}
\end{figure*}

\subsection{K-mouflage}

For the K-mouflage models, the scalar energy density is given by \cite{Brax:2014wla}
\be
\rho_\phi= M^4( -K(\bar\chi) +2\bar\chi K'(\bar\chi))
\ee
where we denote by $\bar K= K(\bar \chi)$ and $\bar K'= K'(\bar \chi)$. The background value of the reduced kinetic energy $\bar\chi$ is obtained from
the Klein-Gordon equation which gives exactly
\be
\bar\chi= \frac{\beta^2 \rho_m^2 t^2}{2m_{\rm Pl}^2 \bar K'^2 M^4}
\ee
The dynamics can be entirely characterised by the time evolution of the Hubble rate
\be
\frac{dH}{dt}=-\frac{1}{2m_{\rm Pl}^2}( 2 M^4 \bar\chi \bar K'  + A \rho_m)
\ee
where we specify that at $z=0$ we have that $H_J(z=0)=H_0$ and $A(z=0)=1$. This allows us to calculate the time evolution of the Hubble rate and the spectroscopic velocity. We have represented in Figure 6 the redshift dependence of $\Delta v$ for the cubic K-mouflage model with $K_0=1$. For objects at redshift $z\gtrsim 3$, the difference is significant and turns out to be within reach of future experiments (see below for prospects).

\subsection{Vainshtein}

 In a Friedmann-Robertson-Walker background, the equations of motion of the Galileon can be simplified using $x= \phi'/m_{\rm Pl}$
 where a prime denotes $'=d/d\ln a=- d/d\ln (1+z)$ where $a$ is the scale factor. Defining $\bar y=\frac{\phi}{m_{\rm Pl} x_0}$,  $\bar x= x/x_0$ and $\bar H= H/H_0$ where $H$ is the Hubble rate in the Jordan frame,
the cosmological evolution satisfies \cite{Appleby:2011aa}
\bea
\bar x'&=&-\bar x + \frac{\alpha\lambda -\sigma\gamma}{\sigma\beta-\alpha \omega}\label{eq:eom1} \\
\bar y'&=& \bar x \label{eq:eom2}\\
\bar H'&=& -\frac{\lambda}{\sigma} + \frac{\omega}{\sigma}(\frac{\sigma\gamma-\alpha\lambda}{\sigma\beta-\alpha\omega})\label{eq:eom3}\\
\eea
\begin{figure*}
\centering
\epsfxsize=7 cm \epsfysize=7 cm {\epsfbox{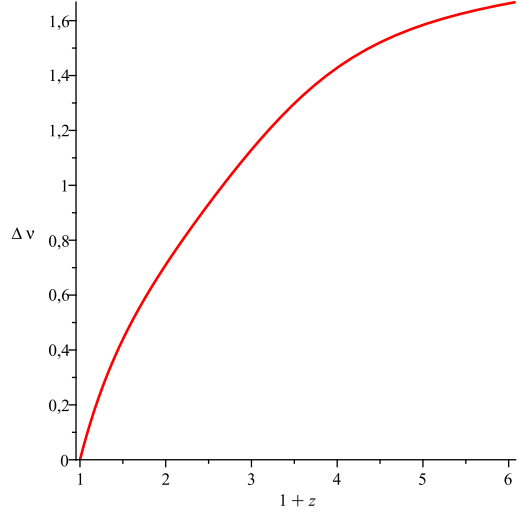}}
\caption{The variation of the spectroscopic velocities $\Delta v$ defined in  equation (3.7)  in cm/s  as a function of  redshift for cubic K-mouflage. }
\end{figure*}
where we have introduced the functions
\begin{align}
\alpha=& -3 \bar c_3 \bar H^3 \bar x^2 +15 \bar c_4 \bar H^5 \bar x^3+\bar \beta \bar H +\frac{\bar c_2 \bar H \bar x}{6} -\frac{35}{2} \bar c_5\bar H^7 \bar x^4  \label{eq:alpha} \\
\beta=& -2 \bar c_3 \bar H^4 \bar x +\frac{\bar c_2\bar H^2}{6} +9 \bar c_4\bar H^6 \bar x^2 -10 \bar c_5 \bar H^8 \bar x^3 \label{eq:beta} \\
\gamma=& 2\bar \beta \bar H^2 -\bar c_3 \bar H^4 \bar x^2 +\frac{\bar c_2 \bar H^2 \bar x}{3} +\frac{5}{2} \bar c_5 \bar H^8 \bar x^4 -2 \bar c_G \bar H^4 \bar x \label{eq:gamma} \\
\sigma=& 2(1-2 \bar \beta \bar y) \bar H -2 \bar \beta \bar H \bar x +2 \bar c_3 \bar H^3 \bar x^3 -15 \bar c_4 \bar H^5 \bar x^4 +21 \bar c_5 \bar H^7 \bar x^5 \label{eq:sigma}\\
\lambda=&  3(1-2 \bar \beta \bar y) \bar H^2 -2\bar \beta\bar H\bar x -2 \bar c_3 \bar H^4 \bar x^3+\frac{\bar c_2 \bar H^2 \bar x^2}{2}+\frac{\Omega_{r0}}{a^4}+\frac{15}{2} \bar c_4\bar H^6 \bar x^4\\
&-9\bar c_5\bar H^8 \bar x^5-\nonumber\\
\label{eq:lambda}\\
\omega=&-2\bar \beta \bar H^2 +2 \bar c_3\bar H^4 \bar x^2-12\bar c_4\bar H^6 \bar x^3+15\bar c_5 \bar H^8 \bar x^4.\label{eq:omega}
\end{align}
The Friedmann equation which governs the evolution of the Hubble rate can be written in a similar way
\begin{equation}
(1-2 \bar \beta \bar y)\bar H^2= \frac{\Omega_{m0}}{a^3}+\frac{\Omega_{r0}}{a^4} +2 \bar \beta \bar H^2 \bar x+\frac{\bar c_2 \bar H^2 \bar x^2}{6}-2\bar c_3 \bar H^4 \bar x^3+\frac{15}{2} \bar c_4 \bar H^6 \bar x^4  -7\bar c_5 \bar H^8 \bar x^5\label{eq:friedman}
\end{equation}
where the final five terms on the right hand side of Equation (\ref{eq:friedman})  correspond to the scalar energy density
\be
\frac{\rho_\phi}{H_0^2m_{\rm Pl}^2}=6 \bar \beta \bar H^2 \bar x+\frac{\bar c_2 \bar H^2 \bar x^2}{2}-6\bar c_3 \bar H^4 \bar x^3+\frac{45}{2} \bar c_4 \bar H^6 \bar x^4
 -21\bar c_5 \bar H^8 \bar x^5-9\bar c_G \bar H^4 \bar x^2.
\ee
The Friedmann equation gives the constraint on the parameters
\be
1= \Omega_{m0}+{\Omega_{r0}} +2 \bar \beta +\frac{\bar c_2}{6}-2\bar c_3 +\frac{15}{2} \bar c_4 -7\bar c_5
\label{con}
\ee
which reduces the dimension of the parameter space by one unit. In the following, we choose $\bar c_2=1$ without any loss of generality implying that
the parameter space comprises $(\bar c_3, \bar c_5, \bar \beta)$ and $\bar c_4$ is determined using (\ref{con}).

\begin{figure}
\centering
\epsfxsize=7 cm \epsfysize=7 cm {\epsfbox{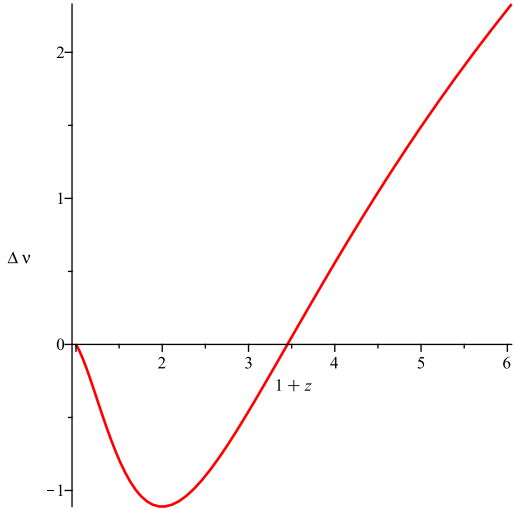}}
\caption{The variation of the spectroscopic velocities $\Delta v$ defined  in  equation (3.7) in cm/s  as a function of  redshift for the quartic Galileon. }
\end{figure}

Numerically, we can adjust the equation of state now to be -1 by choosing for the quartic Galileon, $\bar c_2=1$, $\bar\beta=0.32$, $\bar c_3=1.2$. For these values, we find that the deviation of the spectroscopic velocity deviates from $\Lambda$-CDM in a significant way for objects at redshifts $z\gtrsim 2$. We will discuss how this could be measurable by future observations below.

\subsection{Observational prospects}

The time dependence of the redshift of distant objects (at a redshift $z\gtrsim 2$) can be efficiently probed using absorption lines of the light emitted by distant quasars \footnote{The SKA experiment will probe the redshift drift efficiently for $z\lesssim 1$ but does not have the sensitivity required to distinguish modified gravity models at high redshift \cite{Klockner:2015rqa}.}.
The required precision for these observations, a few cm/s for the spectroscopic velocity, will be attainable  with the E-ELT’s
high-resolution  spectrograph ELT-HIRES. An estimate
of  the spectroscopic velocity precision of such measurements has been given by\cite{Liske:2008ph}
\be
\sigma= 1.35 (\frac{S/N}{2370})^{-1} (\frac{N_{QSO}}{30})^{-1} (1+z)^{-1.7}
\ee
\begin{figure*}
\centering
\epsfxsize=7 cm \epsfysize=7 cm {\epsfbox{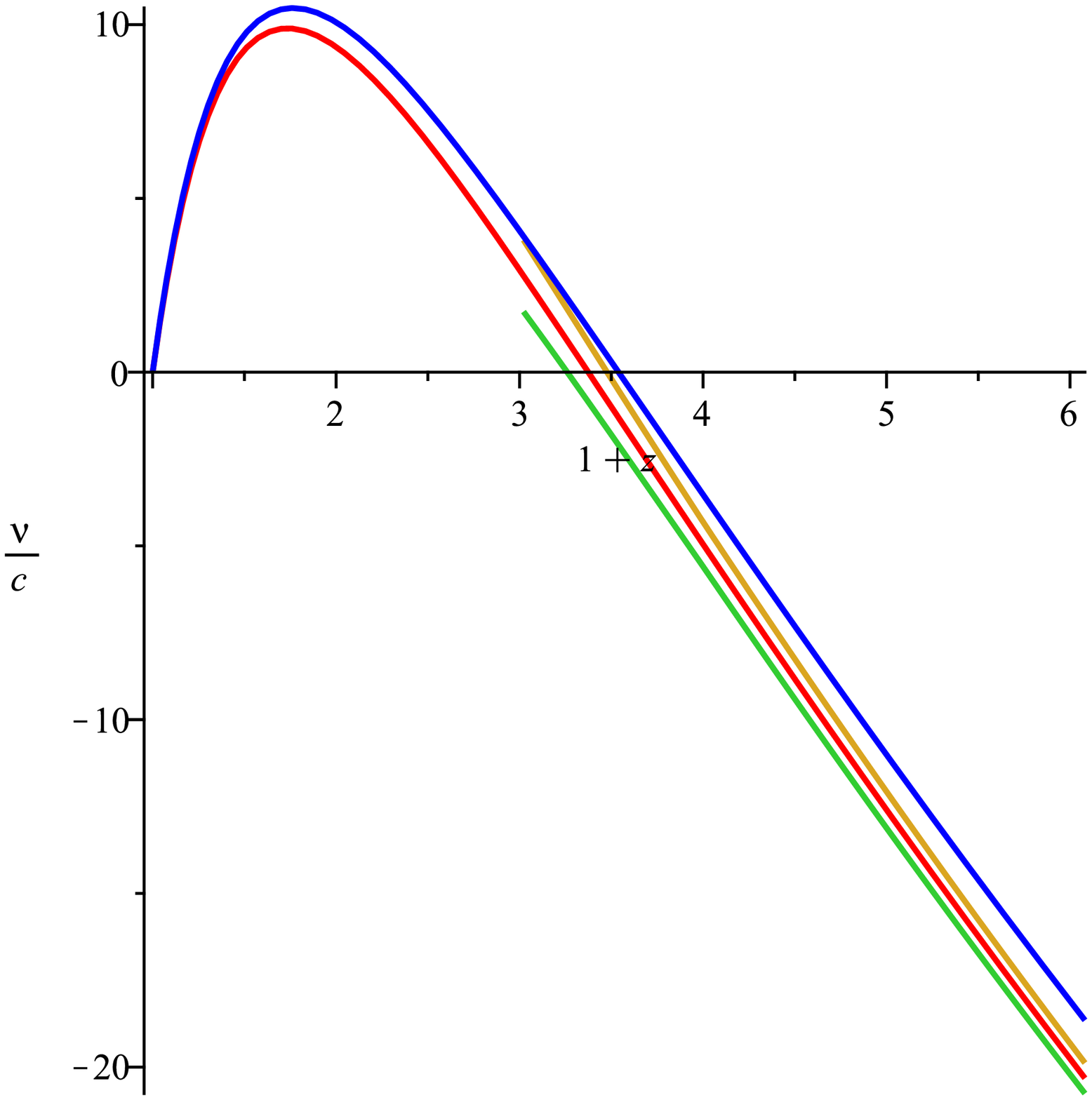}}
\epsfxsize=7 cm \epsfysize=7 cm {\epsfbox{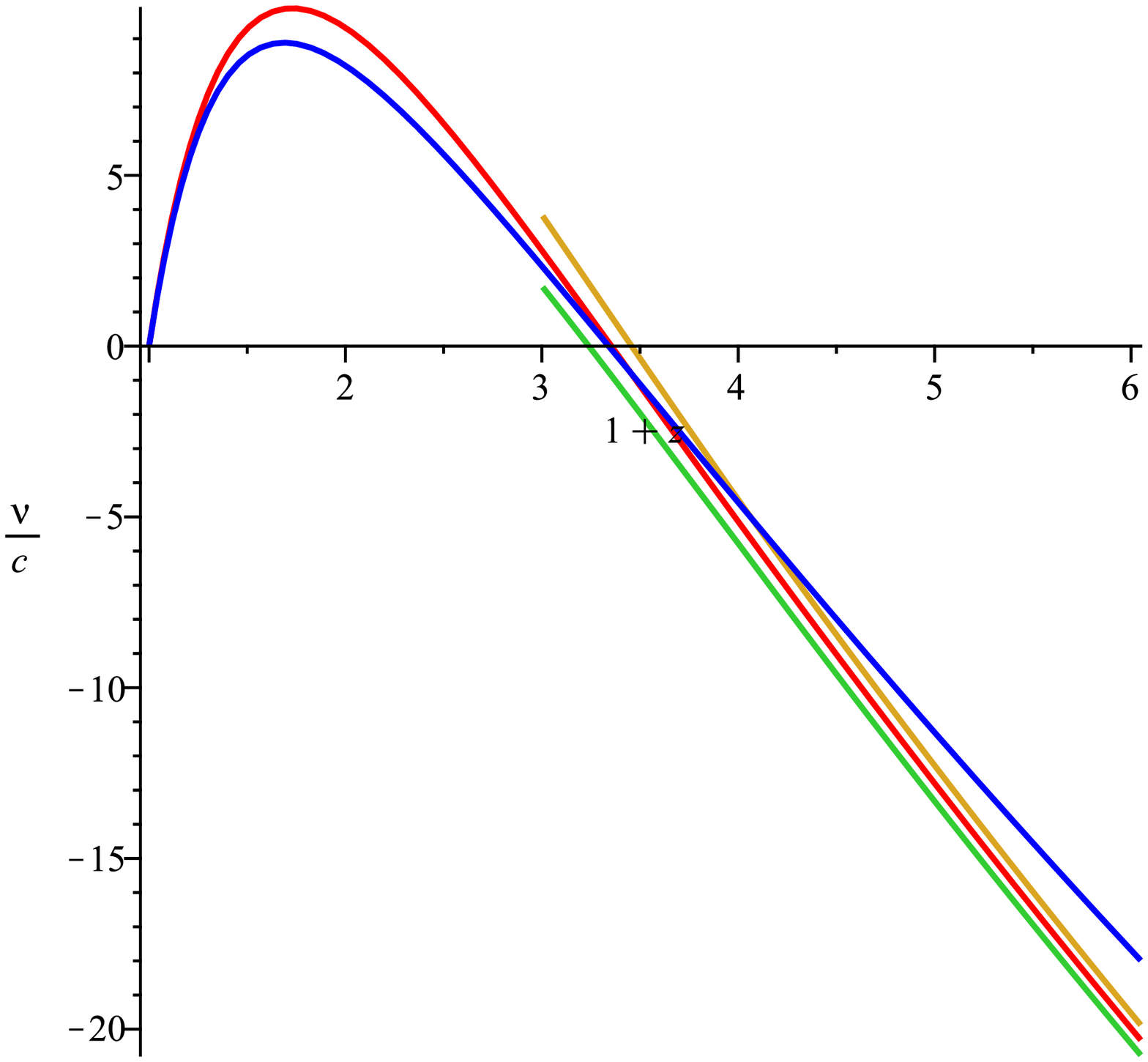}}
\caption{The  spectroscopic velocities $v$ in cm/s for $\Lambda$-CDM (red), the cubic K-mouflage with $K_0=1$ (left-blue) and the quartic Galileon (right-blue) with an equation of state -1 now as a function of  the redshift in the Jordan frame. The expected resolution of  ELT-HIRES around $\Lambda$-CDM lies in the band between the  two curves (green and brown when $z\gtrsim 2$) for  100 quasar absorption systems and a signal to noise ratio of 2000 over 30 years of observation. The K-mouflage and Galileon models deviate by 2$\sigma$ from $\Lambda$-CDM for $z\gtrsim 4$.  }
\end{figure*}
depending on the signal-to-noise of the spectra and  on the number and the redshift of the
quasar absorption systems. The dependence on the redshift has a power -0.9 for $z>4$. In the following, we take a 30 year observation span for $N_{QSO}=100$ systems and a signal to noise ratio of $S/N=2000$. Such a precision, of order a few cm/s' precludes any hope of detecting any effect for chameleon models. On the other hand, both K-mouflage and Galileon models are well within reach. Indeed, in Figure 8, we have plotted
the spectroscopic velocities for the cubic K-mouflage and the quartic Galileon. As a comparison, $\Lambda$-CDM is also displayed as is the expected resolution of ELT-HIRES up to redshifts of $z\sim 5$. In Figure 9
the ratio of the expected deviation of the spectroscopic velocity to its $\Lambda$-CDM counterpart to the expected precision $\sigma$. Galileons with an equation of state of -1 now would be detectable for distant objects of redshift around $z\sim 5$ at the 2$\sigma$ level. For cubic K-mouflage with $K_0=1$, the detection for $z>3$ would be at the same level.
Coupled quintessence models \cite{Corasaniti:2007bg} also give a positive deviation of the spectroscopic velocity at high redshift but with a lower magnitude. On the contrary, the unscreened runaway dilaton \cite{martins} gives a negative deviation in the same range of redshifts. If we were to choose the same number of quasar absorption systems $N_{QSO}=240$ and the same signal to noise ratio $S/N=3000$ as in \cite{Corasaniti:2007bg}, the deviations of both the cubic K-mouflage and the quartic Galileon models would reach  4$\sigma$  at redshifts $z\gtrsim 4$ as shown in Figure 10.

\begin{figure*}
\centering
\epsfxsize=7 cm \epsfysize=7 cm {\epsfbox{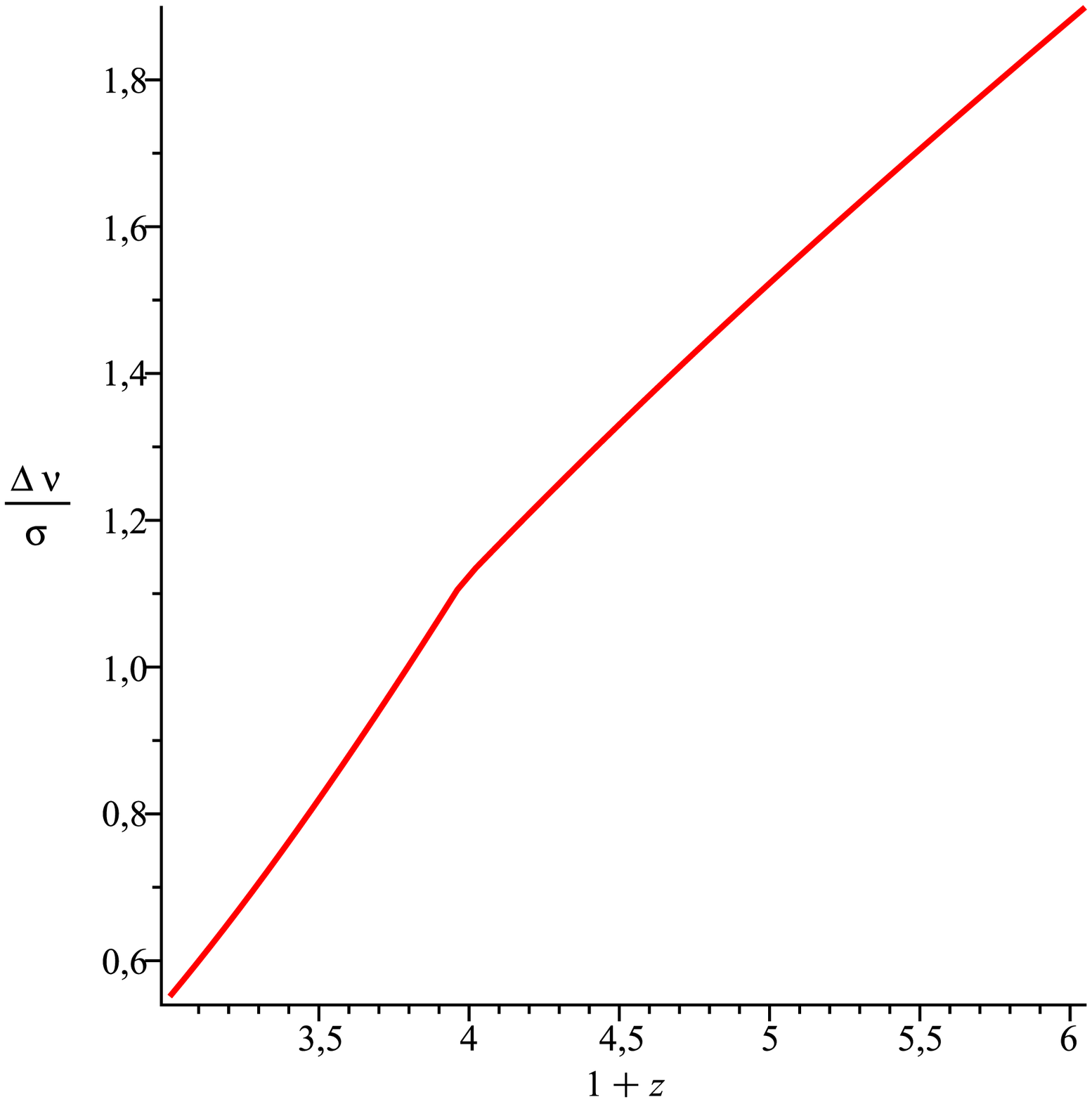}}
\epsfxsize=7 cm \epsfysize=7 cm {\epsfbox{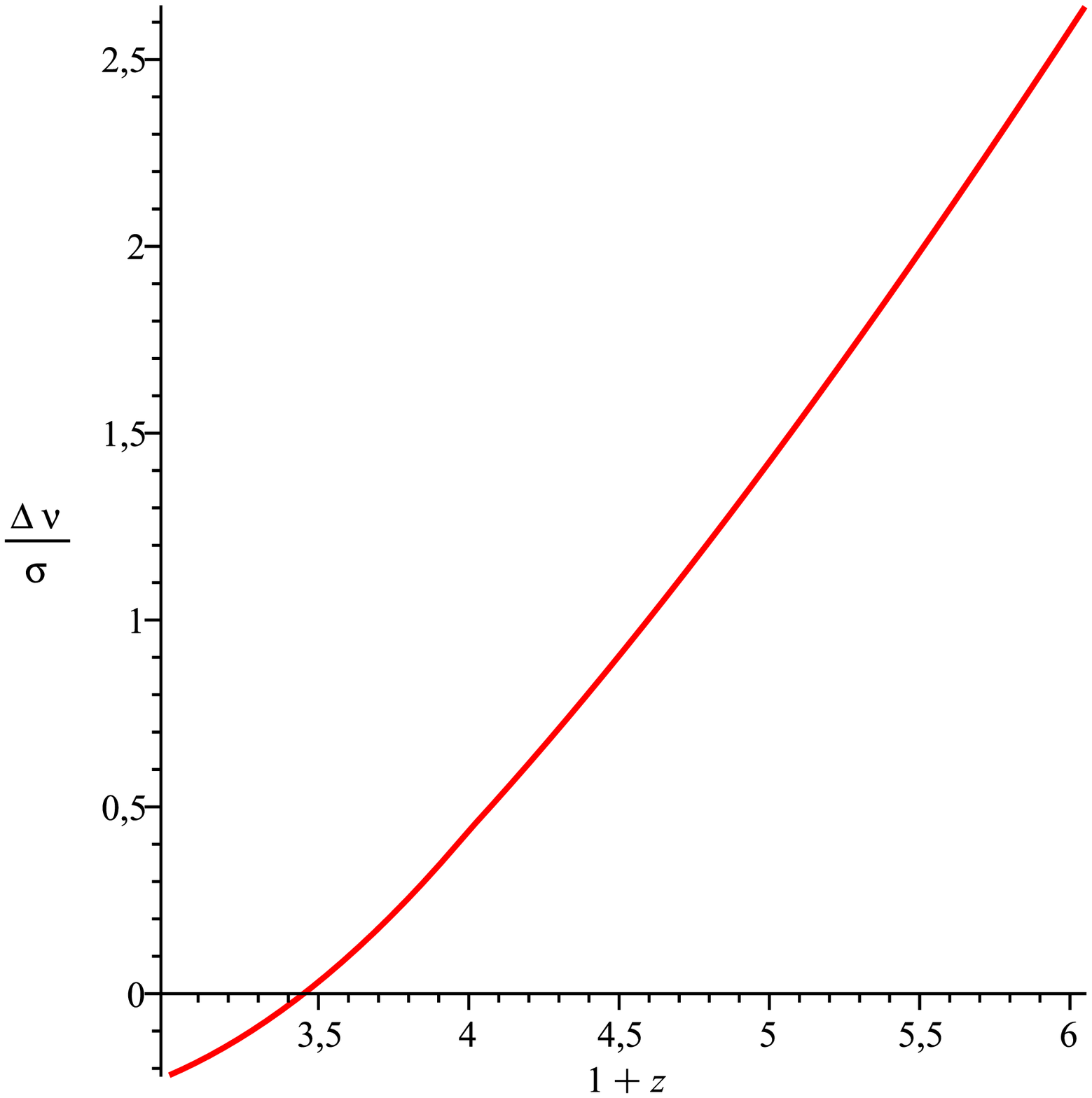}}
\caption{The ratio of the variation of the spectroscopic velocities $\Delta v$ in cm/s compared to the expected precision of future measurements, using the ELT-HIRES with 100 quasar absorption systems and a signal to noise ratio of 2000, as a function of  redshift with  $z\gtrsim 2$  for the cubic K-mouflage model (left) with $K_0=1$ and the quartic Galileon (right) with an equation of state of -1 now. }
\end{figure*}

In summary, we have found that the Sandage effect could become a crucial test for modified gravity models. If a large number of quasar absorption system could be observed, one may even hope that the change of sign of the spectroscopic velocity and the minimum around $z\sim 2$ could be compared to the K-mouflage case with a steady increase in the spectroscopic velocities. Of course, a more thorough investigation of the parameter space of both models should be performed. This is left for future work.

\section{Conclusion}

Modified gravity models fall within three broad categories. In this paper, we have proposed new ways of differentiating them which are not based on effects on the growth of large scale structure. We have shown that chameleon models passing solar system tests can be probed using the variation of the fine structure constant when the coupling of the chameleon to photons is of order one. This is also the case of the proton to mass ratio. In both cases, a clear signal can only be envisaged from unscreened regions of space such as dwarf galaxies. For K-mouflage and Vainshtein, the forthcoming measurements of the time dependence of the redshift of distance objects could be a crucial complement to the study of large scale structure. Indeed their spectroscopic velocity differs from $\Lambda$-CDM significantly for objects at redshifts $z\gtrsim 2$ and we expect that a large class of K-mouflage and Galileon models should be within reach of observations with the HIRES-ELT telescope.

\begin{figure*}
\centering
\epsfxsize=7 cm \epsfysize=7 cm {\epsfbox{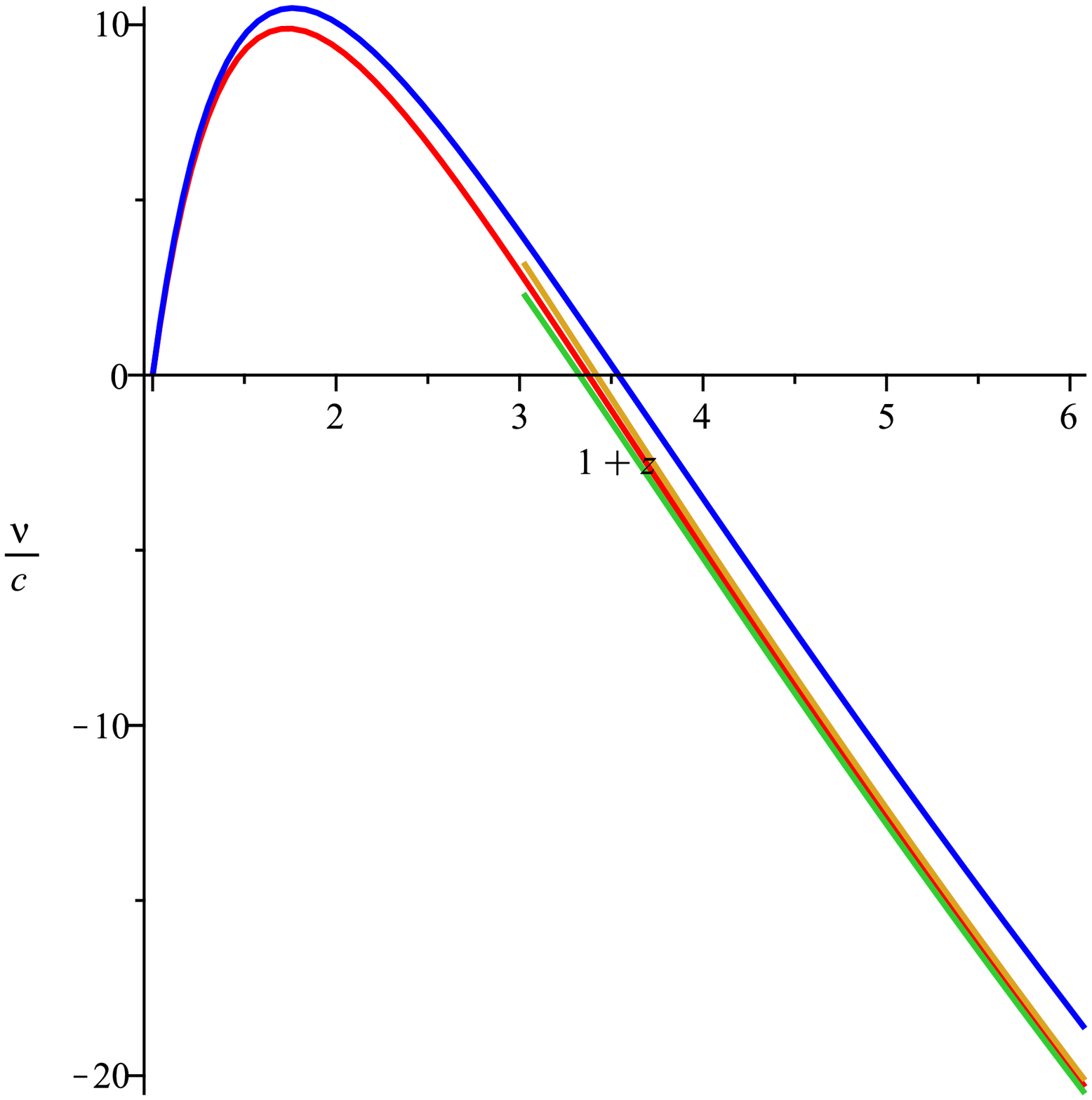}}
\epsfxsize=7 cm \epsfysize=7 cm {\epsfbox{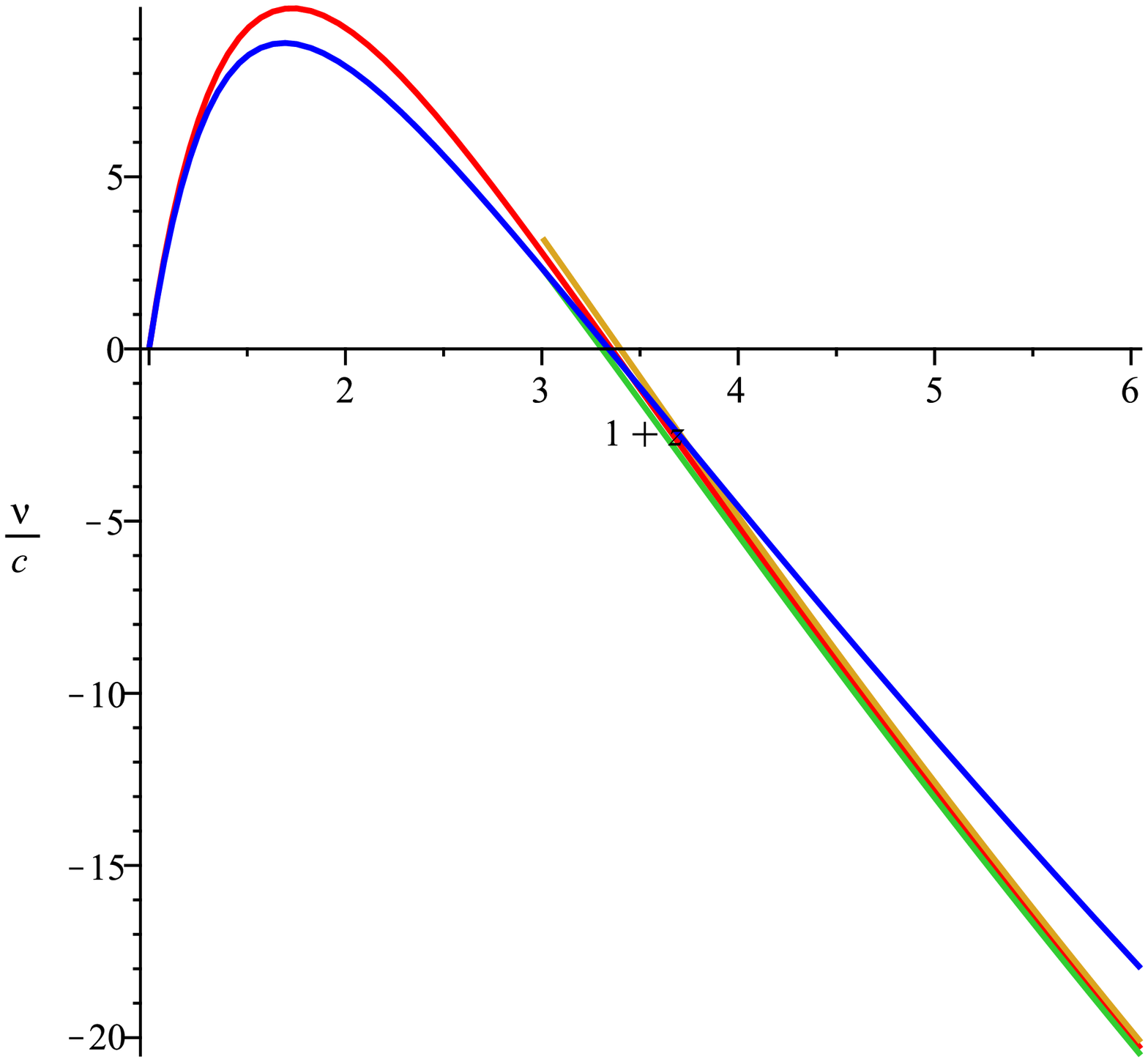}}
\caption{The  spectroscopic velocities $v$ in cm/s for $\Lambda$-CDM (red), the cubic K-mouflage with $K_0=1$ (left-blue) and the quartic Galileon (right-blue) with an equation of state -1 now as a function of  the redshift in the Jordan frame. The expected resolution of  ELT-HIRES around $\Lambda$-CDM lies in the band between the  two curves (green and brown when $z\gtrsim 2$) for  240 quasar absorption systems and a signal to noise ratio of 3000 over 30 years of observation. The K-mouflage and Galileon models deviate by 4$\sigma$ from $\Lambda$-CDM for $z\gtrsim 4$.  }
\end{figure*}

\section{Acknowledgments}

We would like to thank C. Martins, P. Molaro and P. Valageas for suggestions on the manuscript.
P.B.
acknowledges partial support from the European Union FP7 ITN
INVISIBLES (Marie Curie Actions, PITN- GA-2011- 289442) and from the Agence Nationale de la Recherche under contract ANR 2010
BLANC 0413 01. ACD acknowledges partial support from STFC under grants
ST/L000385/1 and ST/L000636/1.

\end{document}